\documentclass[aip,reprint,floatfix,jcp,amsmath]{revtex4-1}
\usepackage{amsfonts}
\usepackage{amsmath}
\usepackage[]{graphicx}
\usepackage{color}
\usepackage{bm}

\draft % marks overfull lines with a black rule on the right

\newcommand{\colr}{}

\newcommand{\NA}{{N_{\rm A}}}

\newcommand{\SL}{{\rm SL}}

\begin{document}

\title{The correlation of the $\NA$ measurements by counting $^{28}$Si atoms}

\author{M.\ Borys}
\affiliation{PTB -- Physikalisch-Technische Bundesanstalt, Bundesallee 100, 38116 Braunschweig, Germany}
\author{I.\ Busch}
\affiliation{PTB -- Physikalisch-Technische Bundesanstalt, Bundesallee 100, 38116 Braunschweig, Germany}
\author{K.\ Fujii}
\affiliation{NMIJ -- National Metrology Institute of Japan, 1-1-1 Umezono, Tsukuba, Ibaraki 305-8563, Japan}
\author{N.\ Kuramoto}
\affiliation{NMIJ -- National Metrology Institute of Japan, 1-1-1 Umezono, Tsukuba, Ibaraki 305-8563, Japan}
\author{G.\ Mana}
\email[]{g.mana@inrim.it}
\affiliation{INRIM -- Istituto Nazionale di Ricerca Metrologica, Str.\ delle Cacce 91, 10135 Torino, Italy}
\author{E.\ Massa}
\affiliation{INRIM -- Istituto Nazionale di Ricerca Metrologica, Str.\ delle Cacce 91, 10135 Torino, Italy}
\author{S.\ Mizushima}
\affiliation{NMIJ -- National Metrology Institute of Japan, 1-1-1 Umezono, Tsukuba, Ibaraki 305-8563, Japan}
\author{T.\ Narukawa}
\affiliation{NMIJ -- National Metrology Institute of Japan, 1-1-1 Umezono, Tsukuba, Ibaraki 305-8563, Japan}
\author{A.\ Nicolaus}
\affiliation{PTB -- Physikalisch-Technische Bundesanstalt, Bundesallee 100, 38116 Braunschweig, Germany}
\author{A.\ Pramann}
\affiliation{PTB -- Physikalisch-Technische Bundesanstalt, Bundesallee 100, 38116 Braunschweig, Germany}
\author{C.\ P.\ Sasso}
\affiliation{INRIM -- Istituto Nazionale di Ricerca Metrologica, Str.\ delle Cacce 91, 10135 Torino, Italy}
\author{M.\ Stock}
\affiliation{BIPM -- Bureau International des Poids et Mesures BIPM, Pavillon de Breteuil, 92312 S\`evres Cedex, France}

\date{\today}

\begin{abstract}
An additional value of the Avogadro constant was obtained by counting the atoms in isotopically enriched Si spheres. With respect to the previous determination, the spheres were etched and repolished to eliminate metal contaminations and to improve the roundness. In addition, all the input quantities -- molar mass, lattice parameter, mass, and volume -- were remeasured aiming at a smaller uncertainty. In order to make the values given in the Refs.\ \onlinecite{Andreas:2011b,Azuma:2015} usable for a least squares adjustment, we report about the estimate of their correlation.
\end{abstract}


\maketitle %\maketitle must follow title, authors, abstract and \pacs

% Constants, fundamental, 06.20.Jr
% Metrology, 06.20.-f
% metrology, 06.20.F-, 06.20.fb (standards) Measurement units and standards, 06.20.F-
% Error theory, 06.20.Dk
% Statistics, 02.50.-r
% 06.20.Dk Measurement and error theory
% 07.05.Kf Data analysis:

\section{Introduction}
In 2011, the International Avogadro Coordination published a comprehensive survey of the measurements contributing to the determination of the Avogadro constant, $\NA$, by counting the atoms in two $^{28}$Si-enriched single crystals -- named AVO28-S5 and AVO28-S8 -- shaped as nearly perfect spheres.\cite{Massa:2011f} This count is a candidate to the realization of the kilogram definition based on fixing the value of the Planck constant.

A limit of the 2011 determination was a metallic contamination -- occurring as metal silicides -- of the sphere surfaces. This contamination was selectively removed by Freckle$^{\rm TM}$ etch;\cite{Nicolaus:2013} subsequently, both spheres were repolished to improve their roundness. Eventually, measurements were repeated by using the repolished spheres -- now, named AVO28-S5c and AVO28-S8c -- and an additional $\NA$ value was published in 2015.\cite{Azuma:2015}

To make both the values usable for the least squares adjustment of the Planck constant value and the kilogram redefinition,\cite{CODATA:2012a,CODATA:2012b} we report about the estimate of their correlation. Section \ref{equation} outlines the measurement procedure; the model used to calculate the correlation of the elemental contributions to the error budgets is given in section \ref{model}. In section \ref{corre:1} we examine in detail the contributions to the $\NA$ uncertainty and iterate the correlation calculation up to the $\NA$ values. The result is given in section \ref{corre:2}.

\section{Measurement equation}\label{equation}
The value of the Avogadro constant,
\begin{equation}\label{NA}
 \NA = \frac{8MV}{a^3 m} ,
\end{equation}
was obtained from measurements of the molar volume, $VM/m$, and lattice parameter, $a$, of two chemically pure $^{28}$Si single-crystals shaped as nearly perfect spheres having about 93 mm diameter. In (\ref{NA}), $m$ and $V$ are the mass and volume, $M$ is the mean molar mass, $a^3/8$ is the atom volume, and 8 is the number of atoms in the cubic unit cell.

From (\ref{NA}), it follows that the $\NA$ determination requires the measurement of i) the lattice parameter -- by combined x-ray and optical interferometry\cite{Massa:2011a,Massa:2015}, ii) the amount of substance fraction of the Si isotopes and, then, of the molar mass -- by absolute mass-spectrometry\cite{Pramann:2011,Narukawa:2014,Vocke:2014,Pramann:2014}, and iii) the sphere mass and volume.\cite{Picard:2011,Kuramoto:2011,Bartl:2011}

Silicon crystals contain impurities, interstitial atoms, and vacancies, which implies that the measured mass value does not correspond to that of an ideal Si crystal and that the crystal lattice may be distorted. This means that the spheres must be characterized both structurally and chemically, so that the appropriate corrections are applied.\cite{Fujimoto:2011,Massa:2011b,Zakel:2011,Melis:2015} The mass, thickness and chemical composition of the surface layers must be taken into account; they are measured by optical and x-ray spectroscopy and reflectometry.\cite{Busch:2011}

\section{Correlation model}\label{model}
The calculation of the correlation of the 2011 and 2015 results requires the correlation of each pair of input data -- molar mass, lattice parameter, volume, and mass. In turn, it is necessary to estimate the correlations of all the elemental contributions to the uncertainties of the 2011 and 2015 input values.

To exemplify the estimation of these correlations, let us consider an elemental contribution to the total uncertainty of a given input datum. We indicate by $x_{11}$ and $x_{15}$ the 2011 and 2015 estimates -- which might be zero -- of the relevant corrections $y$ and $z$. Hence,
\begin{subequations} \begin{eqnarray}
 x_{11} &=&y + u_{11} \\
 x_{15} &=&z + u_{15} ,
\end{eqnarray} \end{subequations}
where $u_{11}$ and $u_{15}$ are zero-mean errors having variances $\sigma_{u11}^2$ and $\sigma_{u15}^2$. Each error is split in two parts, that is,
\begin{subequations} \begin{eqnarray}
 u_{11} &=&\kappa_{11}\epsilon_0 + \epsilon_{11} \\
 u_{15} &=&\kappa_{15}\epsilon_0 + \epsilon_{15}
\end{eqnarray} \end{subequations}
{\colr where $\epsilon_0$ is a zero-mean systematic error affecting identically both the estimates,} $\kappa_{11}$ and $\kappa_{15}$ are scale factors, and $\epsilon_{11}$ and $\epsilon_{15}$ are zero-mean random errors. All these errors are assumed to be uncorrelated. In matrix notation, $\mathbf{u} = \mathbb{W} \boldsymbol{\epsilon}$, where $\mathbf{u} = [u_{11}, u_{15}]^T$, $\boldsymbol{\epsilon} = [\epsilon_0, \epsilon_{11}, \epsilon_{15}]^T$, and the design matrix
\begin{equation}
 \mathbb{W} = \left( \begin{array}{ccc}
    \kappa_{11} & 1 & 0 \\
    \kappa_{15} & 0 & 1 \\
  \end{array} \right)
\end{equation}
maps the $\boldsymbol{\epsilon}$ into $\mathbf{u}$. Since the $\boldsymbol{\epsilon}$ contributions to the total errors $\mathbf{u}$ are uncorrelated, their covariance matrix is
\begin{equation}
 C_{\epsilon\epsilon} = \left( \begin{array}{ccc}
    \sigma_0^2 & 0 & 0 \\
    0 & \sigma_{11}^2 & 0 \\
    0 & 0 & \sigma_{15}^2 \\
  \end{array} \right) ,
\end{equation}
where $\sigma_0^2$ is the variance of the systematic error and $\sigma_{11}^2$ and $\sigma_{15}^2$ are the variances of the random errors. Eventually, the covariance matrix of $\mathbf{u}$,
\begin{equation}
 C_{uu} = \mathbb{W} C_{\epsilon\epsilon} \mathbb{W}^T = \left( \begin{array}{cc}
    \kappa_{11}^2\sigma_0^2+\sigma_{11}^2 & \kappa_{11}\kappa_{15}\sigma_0^2  \\
    \kappa_{11}\kappa_{15}\sigma_0^2 & \kappa_{15}^2\sigma_0^2+\sigma_{15}^2  \\
  \end{array} \right),
\end{equation}
is given by the law of propagation of the uncertainty. It is worth noting that the covariance matrices of $\{u_{11}, u_{15}\}$ and $\{x_{11}, x_{15}\}$ are the same. Therefore, we can refer either to the correlation of the applied corrections or of the correction errors.

The correlation of $u_{11}$ and $u_{15}$ is the ratio between $\kappa_{11}\kappa_{15}\sigma_0^2$ and the geometric mean $\widetilde{\sigma}_u^2 = \sigma_{u11}\sigma_{u15}$ of the variances $\sigma_{u11}^2=\kappa_{11}^2\sigma_0^2+\sigma_{11}^2$ and $\sigma_{u15}^2=\kappa_{15}^2\sigma_0^2+\sigma_{15}^2$ of $u_{11}$ and $u_{15}$, that is, $\rho = \kappa_{11}\kappa_{15}\sigma_0^2/\widetilde{\sigma}_u^2$. {\colr The $\kappa_{11}\kappa_{15}$ sign determines if the elemental uncertainty-contributions are correlated, $\kappa_{11}\kappa_{15} > 0$, or anti-correlated, $\kappa_{11}\kappa_{15} < 0$.

As a guideline, it can be noted that the correlation,
\begin{equation}\label{rho}
 \rho = \frac{\kappa_{11}\sigma_0}{\sigma_{11}} \frac{\kappa_{15}\sigma_0}{\sigma_{15}}
\end{equation}
is the product of the systematic fractions of the total uncertainties, $\kappa_{11}\sigma_0/\sigma_{11}$ and $\kappa_{15}\sigma_0/\sigma_{15}$.} Two limit cases are worth consideration. The first is when the same systematic error affects both the correction estimates, that is, when $\kappa_{11}=\kappa_{15}=1$. In this case, $\rho = \sigma_0^2/\widetilde{\sigma}_u^2$. The second is when the scales of the systematic error are proportional to the standard deviations of the total errors, that is, when $\kappa_{11} \rightarrow \kappa_{11}\sigma_{u11}/\sigma_0$ and $\kappa_{15} \rightarrow \kappa_{15}\sigma_{u15}/\sigma_0$. In this case, $\rho = \kappa_{11}\kappa_{15}$.

\section{Correlation of the input quantities}\label{corre:1}
In the following we examine the uncertainties and correlations of the 2011 and 2015 values of the quantities -- molar mass, lattice parameter, volume, and mass -- measured to determine $\NA$. These quantities were separately measured for each sphere and the two $\NA$ values thus obtained were subsequently averaged. In addition, at least two laboratories measured the input quantities, but the lattice parameter, and the results were averaged. To simplify the analysis, we consider only the $\NA$ values obtained by using AVO28-S5 and AVO28-S5c. It is worth noting that, since these values do not benefit of the averaging, the associated uncertainties are a bit larger than those given in Refs.\ \onlinecite{Andreas:2011b,Azuma:2015}.

To estimate the correlation of the 2011 and 2015 values of the input quantities, we split each elemental contribution to their total uncertainties in systematic and random parts. Next, we calculate each correlation as the product of the 2011 and 2015 systematic fractions.

\subsection{molar mass}
The molar mass,
\begin{equation}\label{molar}
 M = M_{28} + (M_{29}-M_{28})x_{29} + (M_{30}-M_{28})x_{30} ,
\end{equation}
is given in terms of the molar masses $M_i$ and amount-of-substance fractions $x_i$ of the Si isotopes. For the 2015 determination, the amount-of-substance fractions were remeasured independently and {\it ab initio} by the PTB, NMIJ, and National Institute of Standards and Technologies, using isotope dilution\cite{Mana:2010} and multicollector inductively coupled plasma mass spectrometers\cite{Narukawa:2014,Vocke:2014,Pramann:2014}, and by the INRIM, via neutron activation.\cite{D'Agostino:2015} Contrary to the 2011 measurement, tetramethylammonium hydroxide was used as solvent and diluent to reduce the baseline level of the ion currents in the mass spectrometry. The values of the molar mass of the Si isotopes are extremely well known and contributed negligibly to the total uncertainties. Therefore, the 2011 and 2015 values of the molar values are uncorrelated.

\begin{table}
\caption{\label{table:d220} Relative uncertainties (expressed in parts per $10^9$) and correlations of the elemental contributions to the uncertainties of the determinations of the AVO28-S5 and AVO28-S5c lattice parameter. The data are from the tables 1 and I of Refs.\ \onlinecite{Massa:2011a} and \onlinecite{Massa:2015}. The extrapolation uncertainty is from the Appendix C of Ref.\ \onlinecite{Andreas:2011b}.}
\begin{ruledtabular}
\begin{tabular}{lrrr}
                         &AVO28-S5    &AVO28-S5c \\
contribution             &uncertainty &uncertainty &correlation \\
data averaging           & 0.36       & 0.722 &0.00 \\
wavelength               & 0.03       & 0.033 &0.81 \\
laser beam diffraction   & 0.65       & 0.597 &0.25 \\
laser beam alignment     & 0.77       & 0.480 &0.00 \\
beam walks               & 1.20       & 0.577 &0.00 \\
Abbe's errors            & 1.50       & 0.611 &0.00 \\
movement direction       & 0.65       & 0.214 &0.22 \\
temperature              & 2.55       & 0.497 &0.05 \\
thermal strain           & --         & 0.641 &--   \\
self-weigh               & 0.30       & 0.377 &0.00 \\
aberrations$^a$          & 0.64       & 0.642 &1.00 \\
extrapolation            & 0.70       & 0.697 &1.00 \\
total                    & 3.57       & 1.89  &0.15 \\
\end{tabular}
\begin{flushleft}
$^a$the 2011 uncertainty has been updated to account for the surface stress
\end{flushleft}
\end{ruledtabular}
\end{table}

\subsection{lattice parameter}\label{s:d220}
We determined the lattice parameter midway the spheres AVO28-S5 and AVO28-S8 by measuring -- via a combined x-ray and optical interferometer -- the same crystal displacement in terms of the periods of the travelling x-ray and optical fringes.\cite{Massa:2011a,Massa:2015} Since the point defects (i.e., impurity atoms and vacancies) strain the crystal lattice, the AVO28-S5 and AVO28-S5c lattice-parameter,
\begin{equation}\label{parameter}
 a = a_0 \left[ 1 + \sum_i \beta_i (N_i - N_{0,i}) \right] ,
\end{equation}
was obtained -- in 2011 and 2015, respectively -- by extrapolating the value $a_0$ measured by x-ray/optical interferometry to the spheres' position.\cite{Massa:2011b} In (\ref{parameter}), $N_i$ and $N_{0,i}$ are the concentrations of the i-th defect in the x-ray interferometer and spheres and $\beta_i$ is the relevant sensitivity coefficient. The table \ref{table:d220} summarizes the uncertainties and correlations of the elemental contributions to the uncertainties of the 2011 and 2015 determinations.

The x-ray/optical interferometer used in 2011 was disassembled, upgraded, and reassembled. Therefore, the data averages are assumed uncorrelated.

In 2015, a 532 nm frequency-doubled Nd:YAG laser substituted for the 633 nm diode laser. The wavelength is affected by the pressure of the residual gas in the vacuum chamber. Since none of the chamber, pumps, and pressure gauge were changed, the systematic fractions of the wavelength errors are both estimated as 0.9, to which a 0.81 correlation will correspond.

In 2015, the optical interferometer, the chamber windows, and the delivery, collimation, modulation, pointing, and detection systems of the laser beam were rebuilt or replaced. Therefore, the errors due to beam wanderings synchronous with the displacements of the x-ray interferometer are uncorrelated.

The models and ancillary measurements required to cope with diffraction did not yet pass an experimental verification. {\colr Since we cannot exclude imperfections in the modelling and measurements, we assign to the systematic fractions of the 2011 and 2015 total uncertainties, $\kappa_{11}\sigma_0/\sigma_{11}$ and $\kappa_{15}\sigma_0/\sigma_{15}$ in (\ref{rho}), a uniform probability in the $[0,1]$ interval. Therefore, $\rho \sim -{\rm If}(0<z<1)\ln(z)$ and the expected correlation is 0.25.}

{\colr Since in 2015 the laser beam was aligned by using a different procedure and the estimate of the relevant correction was refined, the alignment errors are not correlated.}

The offsets between the centroids of the x-ray and laser-beam spots were independently nullified. Consequently, the correlation of the Abbe errors is zero.

The error due to the projections of the crystal displacements on the normals to the front mirror and diffracting planes is the baseline misalignment $\alpha \approx 10(2)$ $\mu$rad -- which is common to both the measurements -- times the angular distance $\beta \approx 100(10)$ $\mu$rad (2011) or $\beta \approx 70(10)$ $\mu$rad (2015) of the movement direction from the normals' bisecting-plane. Consequently, the systematic fractions and correlation of the correction errors are equal to 1/3, 2/3, and 0.22, respectively.

As regards the temperature measurements, we used the same thermometer and fixed-point cells in both the 2011 and 2015 measurements, but the thermometer was calibrated anew and read by means of a new resistance bridge. Correlation arises because of the cell offsets and deviations of the thermometer reading from the interpolation function, which together contribute to the uncertainty by about 0.11 mK. Since the total measurement uncertainties are about 1 mK and 0.2 mK, the systematic fractions and correlation of the measurement errors are 1/10, 1/2, and 0.05.

The self-weigh strain of the x-ray interferometer depends on the distance of the support points; since we do not envisage any correlation between these distances and between the strain estimates, the strain corrections are not correlated.

Since we always used the same crystal, the interferometer aberrations effected the measurements in the same way and are fully correlated. In 2015, the associated uncertainty was better evaluated and an additional contribution due to surface stress was included\cite{Quagliotti:2013}; therefore, the 2011 contribution in table \ref{table:d220} has been updated.

The lattice parameters of AVO28-S5 and AVO28-S5c were extrapolated by taking the gradients of the point-defect concentrations between the spheres and the interferometer into account. Since, apart the tiniest effect of the newly determined gradient of the nitrogen concentration, the same extrapolations were carried out in 2011 and 2015, they are fully correlated.

{\colr By using a matrix notation, the total errors, $\mathbf{u}_a = [u_{a,11}, u_{a,15}]^{\rm T}$, affecting the 2011 and 2015 measurements of the lattice parameter are
\begin{equation}
 \mathbf{u}_a = \mathbb{A} \mathbf{u} ,
\end{equation}
where $\mathbf{u}=[u_{1,11},\, ...\, u_{11,11}, u_{1,15},\, ...\, u_{12,15}]^{\rm T}$ are the elemental terms listed in table \ref{table:d220} and the $2\times 23$ design matrix is
\begin{equation}
 \mathbb{A} = \left( \begin{array}{cccccc}
           1 & ... & 1 & 0 & ... & 0 \\
           0 & ... & 0 & 1 & ... & 1 \\
         \end{array} \right) .
\end{equation}
After writing the $23 \times 23$ covariance matrix $C_{uu}$ of $\mathbf{u}$ by using the uncertainties and correlations listed in table \ref{table:d220}, the matrix of the relative variances and covariances of the lattice parameter values is
\begin{equation}\label{Cd}
 C_{aa} = \mathbb{A} C_{uu} \mathbb{A}^T = \left( \begin{array}{cc}
 12.8  & 1.03 \\
 1.03  & 3.56 \\
 \end{array} \right) \times 10^{-18} ,
\end{equation}
yielding a correlation of 15\%. Actually, the equation $\mathbb{A} C_{uu} \mathbb{A}^T$ holds only if $C_{uu}$ contains the absolute variances and covariances. In (\ref{Cd}), the use of relative quantities is possible because the two measured values of $\NA$ are extremely close.}

\begin{table}
\caption{\label{table:volumeNMIJ} Relative uncertainties (expressed in parts per $10^9$) and correlations of the elemental contributions to the uncertainties of the determinations of the volumes of the Si-cores of AVO28-S5 and AVO28-S5c at the NMIJ. The data are from tables 4 and 5 of Refs.\ \onlinecite{Kuramoto:2011} and \onlinecite{Azuma:2015}.}
\begin{ruledtabular}
\begin{tabular}{lrrr}
                          &AVO28-S5   &AVO28-S5c \\
                          &uncertainty&uncertainty &correlation \\
interferogram analysis    &36.3       & 10.3  &0.00 \\
temperature               & 4.8       & 4.8   &0.21 \\
diffraction               &16.0       &16.0   &0.25 \\
surface$^a$               & 2.5       & 0.8   &0.52 \\
volume estimation         &29.7       & 3.3   &0.00 \\
total                     &49.9       &19.9   &0.07
\end{tabular}
\end{ruledtabular}
\begin{flushleft}
$^a$optical effect $2(\Delta R - t_\SL)$ of the surface
\end{flushleft}
\end{table}

\subsection{volume}\label{s:volume}
The volume of the Si sphere-core, $V=\pi D^3/6$, is given in terms of the mean diameter
\begin{equation}\label{volume}
 D = D_0 + 2(\Delta R - t_\SL) ,
\end{equation}
where $D_0$ is the mean optical diameter, the $\Delta R$ correction takes the phase shift of the laser-beam reflection into account, $D_0 + 2\Delta R$ is the mechanical diameter, and $t_{\SL}$ is the thickness of the surface layers.

Optical interferometers were used to measure the diameters. The measurements take advantage of a differential approach: firstly it is measured the spacing of an optical cavity; secondly, the sphere is placed into the cavity and the gaps are measured. Diameters, measured in many different directions, are obtained by difference. The tables \ref{table:volumeNMIJ} and \ref{table:volumePTB} give the uncertainties and correlations of the elemental contributions to the uncertainties of the 2011 and 2015 measurements.

The optical values $D_0$ of the sphere diameters are affected by a number of errors -- e.g., in the analysis of the interference pattern, laser frequency measurement, phase-step setting, optical alignments. Since the measurements were repeated anew, they are not correlated.

The temperature measurements were carried out by using the same thermometers and fixed-point cells, but the thermometers were calibrated anew. Correlation arises because of the cell offsets and deviations of the thermometer from the interpolation function, which together contribute to the uncertainty by about 0.28 mK. By using the values in table \ref{table:volumeNMIJ} as an example, since to total measurement uncertainties are about 0.62 mK, we estimate that the systematic fractions and correlation of the measurement errors are 0.45, 0.45, and 0.21.

Wavefront distortions and parasitic interferences affect the optical diameters. The former relates to effects as retrace errors due to imperfections of both the sphere and the reference etalons, to misalignments, and to diffraction. The latter relates to interferences between the stray light reflected by the surfaces of the optical system. This problem affected the 2011 measurement of the NMIJ, but it was eliminated by redesigning the interferometer etalons. Also because the poor understanding and modelling of the relevant phenomena, it is difficult to quantify the variance of the relevant errors. {\colr Though the different interferometers and setups used by the NMIJ and PTB and the newly shaped spheres might suggest a relatively low correlation, we cautiously assume again that the systematic fractions of the total uncertainties, $\kappa_{11}\sigma_0/\sigma_{11}$ and $\kappa_{15}\sigma_0/\sigma_{15}$ in (\ref{rho}), have a uniform probability in the $[0,1]$ interval. Therefore, as in section \ref{s:d220}, the expected correlation is 0.25.}

\begin{table}
\caption{\label{table:volumePTB} Relative uncertainties (expressed in parts per $10^9$) and correlations of the elemental contributions to the uncertainties of the determinations of the volumes of the Si-cores of AVO28-S5 and AVO28-S5c at the PTB. The data are from the tables 1 and 6 of Refs.\ \onlinecite{Bartl:2011} and \onlinecite{Azuma:2015}.}
\begin{ruledtabular}
\begin{tabular}{lrrr}
                             &AVO28-S5         &AVO28-S5c \\
                             &uncertainty   &uncertainty &correlation \\
interferogram analysis       & 4.0          & 4.0   &0.00 \\
temperature                  & 6.0          & 6.0   &0.13 \\
diffraction                  &35.0          &25.0   &0.25 \\
surface$^a$                  & 4.8          & 0.5   &0.52 \\
volume estimation            & 6.0          & 6.0   &0.00 \\
total                        &36.6          &26.7   &0.23
\end{tabular}
\begin{flushleft}
$^a$optical effect $2(\Delta R - t_\SL)$ of the surface
\end{flushleft}
\end{ruledtabular}
\end{table}

The optical value of the mean diameter is corrected for the optical effect $2(\Delta R - t_\SL)$ of the surface -- modeled as a layered structure of SiO$_2$, metal contaminants, chemisorbed water, and hydrocarbons. This correction combines the defective optical thickness $\Delta R$ -- due to the reflection phase-shift -- and the total geometrical thickness of the surface layers $t_\SL$. These corrective terms are strongly correlated -- because both depend on the surface model and the thicknesses of each model layer -- and cancel nearly exactly, as well as their estimate uncertainties. The largest sources of uncertainty are the chemisorbed water and carbonaceous and (only in 2011) metallic contaminations. In turn, the uncertainties of these layers affect also the estimate of the SiO$_2$ thickness. The 2011 and 2015 errors of the thicknesses of the chemisorbed water and carbonaceous contaminations are strongly correlated because the former thicknesses were estimated from the same literature data and the latter were measured by using same reference. Since, before the 2015 measurements, the metallic contamination was eliminated and the sphere surface re-oxidized, we assume that the 2011 and 2015 model errors did not correlate the estimates of the SiO$_2$ thicknesses. Additionally, we assume that the systematic fractions of the $2(\Delta R - t_\SL)$ errors are the same 0.61 and 0.85 systematic-fractions of the $t_\SL$ errors calculated from the tables 10 and 4a in Refs.\ \onlinecite{Busch:2011} and \onlinecite{Azuma:2015}. Accordingly, the correlation of the $2(\Delta R - t_\SL)$ errors is 0.52.

The coverage of the diameter measurements, positioning errors, and goodness of fit affect the volume calculation. Since, owing to the repolishing, the surface geometries differed substantially, the last entries of the tables \ref{table:volumeNMIJ} and \ref{table:volumePTB} are uncorrelated.

As an example, the matrix of the relative variances and covariances of the NMIJ's volume values is
\begin{equation}
 C_{VV} = \left( \begin{array}{cc}
 2485.1  & 69.9 \\
   69.9  & 396.7 \\
 \end{array} \right) \times 10^{-18} ,
\end{equation}
yielding a correlation of 7\%.

{\colr The matrix of the relative variances and covariances of the 2011 and 2015 weighted averages of the NMIJ and PTB values is
\begin{equation}
 C_{VV} = \mathbb{A} \Phi \mathbb{A}^T = \left( \begin{array}{cc}
 995..3  & 73.7 \\
   73.7  & 277.5 \\
 \end{array} \right) \times 10^{-18} ,
\end{equation}
where
\begin{equation}
 \mathbb{A} = \left(
         \begin{array}{cccc}
           1 & 1 & 0 & 0  \\
           0 & 0 & 1 & 1  \\
         \end{array}
       \right)
\end{equation}
is the design matrix of the weighted average of the $[V^{\rm NMIJ}_{11}, V^{\rm PTB}_{11},V^{\rm NMIJ}_{15},V^{\rm PTB}_{15}]^{\rm T}$ list of the input values and, by neglecting the correlation of the NMIJ and PTB estimates of the optical effect of the surface,
\begin{equation}
 \Phi = \left( \begin{array}{cccc}
 2485.1  & 0      & 69.9  & 0 \\
 0       & 1336.0 & 0     &224.7 \\
 69.9    & 0      & 396.7 & 0 \\
 0       & 224.7  & 0     & 713.3 \\
 \end{array} \right) \times 10^{-18} ,
\end{equation}
is the joint matrix of the relative variances and covariances of the 2011 and 2015 NMIJ and PTB volume-values. Eventually, the uncertainty and correlation of the 2011 and 2015 weighted averages of the NMIJ and PTB values are given in table \ref{table:volume}.}

\begin{table}
\caption{\label{table:volume} Uncertainty -- expressed in parts per $10^9$ -- and correlation of the weighted averages of the NMIJ and PTB determinations of the volumes of the Si-cores of AVO28-S5 and AVO28-S5c.}
\begin{ruledtabular}
\begin{tabular}{rrl}
     AVO28-S5    &AVO28-S5c  \\
     uncertainty &uncertainty &correlation \\
      30.9       & 16.7       &0.14
\end{tabular}
\end{ruledtabular}
\end{table}

\subsection{mass}
The sphere masses were determined by comparisons against Pt–Ir or stainless-steel standards both in low pressure argon (by NMIJ in 2011) and in vacuo. The tables \ref{table:mass-NMIJ}, \ref{table:mass-PTB}, and \ref{table:mass-BIPM} give the uncertainties and correlations of the elemental contributions to the uncertainties of the 2011 and 2015 weighings at the NMIJ, PTB, and BIPM.

As an example, the model equations of the mass measurements at the NMIJ are
\begin{subequations} \begin{eqnarray}
 m_{\rm Si}  &= &m_0 + \Delta m \\
 m_0         &= &m_0^{\rm air} + m_{\rm H_2O} \\
 \Delta m    &= &\Delta l/S + \rho(V_{\rm Si} - V_0) + m_0 \gamma \Delta h / g ,
\end{eqnarray} \end{subequations}
where $m_{\rm Si}$ and $m_0$ are the AVO28-S5c and standard masses in vacuum, $\Delta m$ is the observed mass difference, $m_0^{\rm air}$ is the standard mass in air, $m_{\rm H_2O}$ is the mass of the desorbed water, $\Delta l$ is the balance indication, $S$ is the balance sensitivity, $\rho$ is the Ar density (or zero, when weighing in vacuo), $V_{\rm Si}$ and $V_0$ are the AVO28-S5c and standard volumes, $\gamma$ is the gravity vertical-gradient, $g$ is the gravitational acceleration, and $\Delta h$ is the height difference between the centers of mass of AVO28-S5c and the standard.

\begin{table}
\caption{\label{table:mass-NMIJ} Relative uncertainties (expressed in parts per $10^9$) and correlations of the elemental contributions to the uncertainties of the determinations of the masses of the Si-cores of AVO28-S5 and AVO28-S5c at the NMIJ.}
\begin{ruledtabular}
\begin{tabular}{lrrrrr}
                        &AVO28-S5  &AVO28-S5c \\
                        &uncertainty&uncertainty &correlation \\
balance indication      & 3.7      & 5.4   &0.65 \\
washing reproducibility & 4.0      & 4.0   &1.00 \\
air buoyancy            & 0.03     & 0.03  &0.00 \\
reference-mass sorption & 2.1      & 2.1   &1.00 \\
center of mass          & 0.3      & 0.3   &1.00 \\
BIPM mass scale         & 3.0      & 3.0   &1.00 \\
surface layer mass      & 14.5     & 10.0  &0.27 \\
point defect mass       & 2.4      & 3.8   &0.63 \\
total                   & 16.1     & 13.2  &0.41
\end{tabular}
\end{ruledtabular}
\end{table}

\begin{table}
\caption{\label{table:mass-PTB} Relative uncertainties (expressed in parts per $10^9$) and correlations of the elemental contributions to the uncertainties of the determinations of the masses of the Si-cores of AVO28-S5 and AVO28-S5c at the PTB.}
\begin{ruledtabular}
\begin{tabular}{lrrrrr}
                        &AVO28-S5 &AVO28-S5c \\
                        &uncertainty&uncertainty &correlation \\
mass of standards$^a$   & 6.7     & 6.6   &0.23 \\
weighing difference     & 0.5     & 0.5   &0.00 \\
balance linearity       & 0.7     & 0.4   &0.57 \\
balance sensitivity     & 1.7     & 0.2   &0.12 \\
mass position           & 0.5     & 0.5   &0.50 \\
reference-mass sorption & 9.1     & 1.0   &0.00 \\
center of mass          & 0.3     & 0.3   &1.00 \\
surface layer mass      & 14.5    & 10.0  &0.27 \\
point defect mass       & 2.4     & 3.8   &0.63 \\
total                   & 18.6    & 12.6  &0.24
\end{tabular}
\end{ruledtabular}
\begin{flushleft}
$^a$with auxiliary weights and correction for the BIPM mass scale
\end{flushleft}
\end{table}

\begin{table}
\caption{\label{table:mass-BIPM} Relative uncertainties (expressed in parts per $10^9$) and correlations of the elemental contributions to the uncertainties of the determinations of the masses of the Si-cores of AVO28-S5 and AVO28-S5c at the BIPM.}
\begin{ruledtabular}
\begin{tabular}{lrrrrr}
                        &AVO28-S5 &AVO28-S5c \\
                        &uncertainty&uncertainty &correlation \\
weighing A0$^a$         & 1.7     & 1.6   &0.54 \\
weighing$^b$            & 0.9     & 2.7   &0.07 \\
sorption of A0          & 0.6     & 0.8   &0.80 \\
BIPM mass scale         & 3.0     & 3.0   &1.00 \\
surface layer mass      & 14.5    & 10.0  &0.27 \\
point defect mass       & 2.4     & 3.8   &0.63 \\
total                   & 15.1    & 11.6  &0.32
\end{tabular}
\end{ruledtabular}
\begin{flushleft}
$^a$A0 was compared against the reference mass in air\\
$^b$AVO28-S5 and AVO28-S5c were compared against the A0 in vacuo
\end{flushleft}
\end{table}

In 2014, the BIPM carried out a calibration campaign with respect to the international prototype, in anticipation of the planned redefinition of the kilogram.\cite{Stock:2015} This campaign brought to light the existence of an offset in the BIPM as-maintained mass unit, which was traceable to the prototype in 1992. The corrections to apply depend on the date of the BIPM calibrations used for traceability. In the case of NMIJ, both the corrections were $-30.1(3.0)$ $\mu$g; in the case of PTB, the corrections were $-30.4(3.0)$ $\mu$g and $-35.4(3.0)$ $\mu$g. Therefore, the 2011 and 2015 mass values, expressed in the mass unit maintained formerly at the BIPM, were similarly reduced. For the purpose of this analysis, we assume that these corrections are fully correlated.

Such a correlation had always existed: formerly, the BIPM estimated its calibration uncertainty as 7 $\mu$g. This uncertainty was dominated by the lack of knowledge of the mass of the working standards with respect to the international prototype, which was for the last time available in 1992 and since then was stored in a safe. This error correlated all the mass calibrations made by the BIPM. Therefore, the correlation due to the traceability of the mass values to the BIPM is not new and it is not related to the upgrade of the BIPM mass scale.

To obtain the mass of the Si core, the sphere mass is corrected by subtracting the total mass of the surface layers. The SiO$_2$ mass is determined as density times volume; since the same 2.2(1) g/cm$^3$ density was used in the 2011 and 2015 determinations, the density error contributed systematically. The same 0.028(8) $\mu$g/cm$^2$ sorption coefficient of the chemisorbed water was used in 2011 as well in 2015;\cite{Mizushima:2004} therefore, the estimate errors of the chemisorbed-water masses are identical. The masses of the carbonaceous contaminations were determined by comparing the C-K to Si-L peak-ratio against the same reference carbon layer; furthermore, the same CH$_{1.5}$ stoichiometry was assumed. Consequently, the estimated masses of the carbonaceous contaminations are taken fully correlated. Since, in the 2015 measurements, the metallic contamination was removed, it does not contribute to the correlation. Eventually, we extracted the contribution of the SiO$_2$ density, which is systematic, from the total uncertainty of the SiO$_2$ mass and left out the contribution of the SiO$_2$ thickness, which is random. The systematic fractions and correlation of the total-mass errors -- 0.41, 0.66, and 0.27, respectively -- have been obtained from the tables 11 and 4b in Refs.\ \onlinecite{Busch:2011} and \onlinecite{Azuma:2015}.

The point-defect mass was taken into account by correcting the mass values to obtain the mass of a crystal having Si atoms at all lattice sites and no interstitial atoms. The mass deficits -- 8.1(2.4) $\mu$g and 3.8(3.8) $\mu$g for AVO28-S5\cite{Andreas:2011b} and AVO28-S5c,\cite{Azuma:2015} respectively -- were calculated on the basis of the same contaminant and vacancy concentrations, but the 2015 measurement took an additional $-0.3(2)$ $\mu$g correction because of the nitrogen contamination.\cite{Azuma:2015} Furthermore, after the 2011 determination, the spheres AVO28-S5c was thermally oxidised and $-4(3)$ $\mu$g were further subtracted because of a possible metal diffusion into the crystal.\cite{Azuma:2015} Consequently, the systematic fractions and correlation of the 2011 and 2015 estimate errors of the point-defect mass are 1, 0.63, and 0.63, respectively.

As an example, the matrix of the relative variances and covariances of the NMIJ's mass values is
\begin{equation}
 C_{mm} = \left( \begin{array}{cc}
 259.2  & 87.4 \\
 87.4   & 173.1 \\
 \end{array} \right) \times 10^{-18} ,
\end{equation}
yielding a correlation of 41\%.

{\colr The 2011 and 2015 mass values of the NMIJ, PTB and BIPM were averaged by taking the correlations between the individual results into account; details are given in Refs.\ \onlinecite{Azuma:2015} and \onlinecite{Picard:2011}. Next, the same corrections were applied for the masses of the surface layers and point defects. For the sake of simplicity, in table \ref{table:NA}, we use the 32\% correlation of the BIPM values as that of the averaged Si-core masses of AVO28-S5 and AVO28-S5c.}

\begin{table}
\caption{\label{table:NA} Uncertainty budgets -- expressed in parts per $10^9$ -- of the 2011 and 2015 $N_A$ determinations using AVO28-S5 and AVO28-S5c. The uncertainties are from the tables 8 and 12 of Refs.\ \onlinecite{Andreas:2011b} and \onlinecite{Azuma:2015}.}
\begin{ruledtabular}
\begin{tabular}{lrrl}
                       &AVO28-S5    &AVO28-S5c  \\
                       &uncertainty &uncertainty &correlation \\
molar mass             & 7.9     & 5.4   &0.00  \\
unit cell volume       & 10.5    & 5.5   &0.15  \\
sphere volume          & 30.2    & 16.0  &0.14  \\
sphere mass            & 15.0    & 11.0  &0.32  \\
total                  & 36.2    & 20.9  &0.17
\end{tabular}
\end{ruledtabular}
\end{table}

\section{Correlation of the $N_A$ measurements}\label{corre:2}
The table \ref{table:NA} summarizes the uncertainty and correlation of the contributions to the uncertainty of the 2011 and 2015 determinations of $\NA$ using AVO28-S5 and AVO28-S5c. {\colr The correlations are the exemplar values given in the tables \ref{table:d220}, \ref{table:volume}, \ref{table:mass-BIPM}, but the uncertainty contributions are taken from the tables 8 and 12 of Refs.\ \onlinecite{Andreas:2011b} and \onlinecite{Azuma:2015}. The small differences between the uncertainties given in table \ref{table:NA} and in the sections \ref{s:d220} and \ref{s:volume} are due the difficulty to reconstruct the rounding in the calculations.} The joint matrix of the relative variances and covariances is
\begin{equation}
 C_{uu} = \left( \begin{array}{cccccccc}
 62 & 0   & 0   & 0   & 0   & 0 & 0  & 0 \\
 0  & 110 & 0   & 0   & 0   & 9 & 0  & 0 \\
 0  & 0   & 912 & 0   & 0   & 0 & 68 & 0 \\
 0  & 0   & 0   & 225 & 0   & 0 & 0  & 53 \\
 0  & 0   & 0   & 0   & 29  & 0 & 0  & 0  \\
 0  & 9   & 0   & 0   & 0   & 30& 0  & 0  \\
 0  & 0   & 68  & 0   & 0   & 0 & 256& 0 \\
 0  & 0   & 0   & 53  & 0   & 0 & 0  & 121
\end{array} \right) \times 10^{-18} .
\end{equation}
The matrix mapping the contributions in table \ref{table:NA} into the 2011 and 2015 total uncertanties is
\begin{equation}
 \mathbb{A} = \left(
         \begin{array}{cccccccc}
           1 & 1 & 1 & 1 & 0 & 0 & 0 & 0  \\
           0 & 0 & 0 & 0 & 1 & 1 & 1 & 1  \\
         \end{array}
       \right) .
\end{equation}
Eventually, the matrix of the relative variance and covariances of the two $N_A$ values is
\begin{equation}
 C_{\NA\NA} = \mathbb{A} C_{uu} \mathbb{A}^T = \left( \begin{array}{cc}
 1310 & 129 \\
 129  & 436 \\
 \end{array} \right) \times 10^{-18} ,
\end{equation}
to which a 17\% correlation will correspond.

\section{Conclusions}
The forthcoming definition of the kilogram based on a fixed value of the Planck constant is prompting the development of extremely refined technologies and experiments to measure both the Planck and Avogadro constants with the utmost accuracy. In order to gain the maximum benefit from the measurement results, to extract from them all the available information, and to evidence discrepancies, a similarly refined analysis of the measurement uncertainty is desirable. This paper investigated the correlation of the $\NA$ values reported by the International Avogadro Coordination. Despite these values were obtained by counting twice the atoms in the same $^{28}$Si-enriched spheres, owing to the sphere-surface repolishing and the upgrades of many of the measurement technologies, their correlation is 17\%.

{\colr In Ref.\ \onlinecite{Azuma:2015}, the value $\NA= 6.022 140 82(18) \times 10^{23}$ mol$^{-1}$ given in Ref.\ \onlinecite{Andreas:2011b} has been updated to $\NA= 6.02214099(18)\times 10^{23}$ mol$^{-1}$, by taking the recalibration of the of the mass standards into account. The value derived from the new measurements repeated by using the repolished spheres is $\NA= 6.02214076(12)\times 10^{23}$ mol$^{-1}$.\cite{Azuma:2015}} Taking the correlation into account, the average is
\begin{equation}\label{AVO}
 \NA = 6.02214082(11) \times 10^{23}\; {\rm mol}^{-1} ,
\end{equation}
with a relative uncertainty of $1.8\times 10^{-8}$.

To assess the consistency of the $\NA$ given in Refs.\ \onlinecite{Andreas:2011b} and \onlinecite{Azuma:2015}, we observe that their difference is $38(33)\times 10^{-9} \NA$, where we took the 17\% correlation into account in the uncertainty calculation.

\begin{acknowledgments}
This work was jointly funded by the European Me\-trology Research Pro\-gramme (EMRP) par\-ti\-ci\-pa\-ting coun\-tries within the European Association of National Metrology Institutes (EURAMET), the European Union, the Italian ministry of education, university, and research (awarded project P6-2013, implementation of the new SI), and the Grant-in-Aid for Scientific Research (B) (KAKENHI 24360037) from the Japan Society for the Promotion of Science.
\end{acknowledgments}

\bibliography{d220_2014}

\providecommand{\noopsort}[1]{}\providecommand{\singleletter}[1]{#1}%
\begin{thebibliography}{25}%
\makeatletter
\providecommand \@ifxundefined [1]{%
 \@ifx{#1\undefined}
}%
\providecommand \@ifnum [1]{%
 \ifnum #1\expandafter \@firstoftwo
 \else \expandafter \@secondoftwo
 \fi
}%
\providecommand \@ifx [1]{%
 \ifx #1\expandafter \@firstoftwo
 \else \expandafter \@secondoftwo
 \fi
}%
\providecommand \natexlab [1]{#1}%
\providecommand \enquote  [1]{``#1''}%
\providecommand \bibnamefont  [1]{#1}%
\providecommand \bibfnamefont [1]{#1}%
\providecommand \citenamefont [1]{#1}%
\providecommand \href@noop [0]{\@secondoftwo}%
\providecommand \href [0]{\begingroup \@sanitize@url \@href}%
\providecommand \@href[1]{\@@startlink{#1}\@@href}%
\providecommand \@@href[1]{\endgroup#1\@@endlink}%
\providecommand \@sanitize@url [0]{\catcode `\\12\catcode `\$12\catcode
  `\&12\catcode `\#12\catcode `\^12\catcode `\_12\catcode `\%12\relax}%
\providecommand \@@startlink[1]{}%
\providecommand \@@endlink[0]{}%
\providecommand \url  [0]{\begingroup\@sanitize@url \@url }%
\providecommand \@url [1]{\endgroup\@href {#1}{\urlprefix }}%
\providecommand \urlprefix  [0]{URL }%
\providecommand \Eprint [0]{\href }%
\providecommand \doibase [0]{http://dx.doi.org/}%
\providecommand \selectlanguage [0]{\@gobble}%
\providecommand \bibinfo  [0]{\@secondoftwo}%
\providecommand \bibfield  [0]{\@secondoftwo}%
\providecommand \translation [1]{[#1]}%
\providecommand \BibitemOpen [0]{}%
\providecommand \bibitemStop [0]{}%
\providecommand \bibitemNoStop [0]{.\EOS\space}%
\providecommand \EOS [0]{\spacefactor3000\relax}%
\providecommand \BibitemShut  [1]{\csname bibitem#1\endcsname}%
\let\auto@bib@innerbib\@empty
%</preamble>
\bibitem [{\citenamefont {Andreas}\ \emph {et~al.}(2011)\citenamefont
  {Andreas}, \citenamefont {Azuma}, \citenamefont {Bartl}, \citenamefont
  {Becker}, \citenamefont {Bettin}, \citenamefont {Borys}, \citenamefont
  {Busch}, \citenamefont {Fuchs}, \citenamefont {Fujii}, \citenamefont
  {Fujimoto}, \citenamefont {Kessler}, \citenamefont {Krumrey}, \citenamefont
  {Kuetgens}, \citenamefont {Kuramoto}, \citenamefont {Mana}, \citenamefont
  {Massa}, \citenamefont {Mizushima}, \citenamefont {Nicolaus}, \citenamefont
  {Picard}, \citenamefont {Pramann}, \citenamefont {Rienitz}, \citenamefont
  {Schiel}, \citenamefont {Valkiers}, \citenamefont {Waseda},\ and\
  \citenamefont {Zakel}}]{Andreas:2011b}%
  \BibitemOpen
  \bibfield  {author} {\bibinfo {author} {\bibfnamefont {B.}~\bibnamefont
  {Andreas}}, \bibinfo {author} {\bibfnamefont {Y.}~\bibnamefont {Azuma}},
  \bibinfo {author} {\bibfnamefont {G.}~\bibnamefont {Bartl}}, \bibinfo
  {author} {\bibfnamefont {P.}~\bibnamefont {Becker}}, \bibinfo {author}
  {\bibfnamefont {H.}~\bibnamefont {Bettin}}, \bibinfo {author} {\bibfnamefont
  {M.}~\bibnamefont {Borys}}, \bibinfo {author} {\bibfnamefont
  {I.}~\bibnamefont {Busch}}, \bibinfo {author} {\bibfnamefont
  {P.}~\bibnamefont {Fuchs}}, \bibinfo {author} {\bibfnamefont
  {K.}~\bibnamefont {Fujii}}, \bibinfo {author} {\bibfnamefont
  {H.}~\bibnamefont {Fujimoto}}, \bibinfo {author} {\bibfnamefont
  {E.}~\bibnamefont {Kessler}}, \bibinfo {author} {\bibfnamefont
  {M.}~\bibnamefont {Krumrey}}, \bibinfo {author} {\bibfnamefont
  {U.}~\bibnamefont {Kuetgens}}, \bibinfo {author} {\bibfnamefont
  {N.}~\bibnamefont {Kuramoto}}, \bibinfo {author} {\bibfnamefont
  {G.}~\bibnamefont {Mana}}, \bibinfo {author} {\bibfnamefont {E.}~\bibnamefont
  {Massa}}, \bibinfo {author} {\bibfnamefont {S.}~\bibnamefont {Mizushima}},
  \bibinfo {author} {\bibfnamefont {A.}~\bibnamefont {Nicolaus}}, \bibinfo
  {author} {\bibfnamefont {A.}~\bibnamefont {Picard}}, \bibinfo {author}
  {\bibfnamefont {A.}~\bibnamefont {Pramann}}, \bibinfo {author} {\bibfnamefont
  {O.}~\bibnamefont {Rienitz}}, \bibinfo {author} {\bibfnamefont
  {D.}~\bibnamefont {Schiel}}, \bibinfo {author} {\bibfnamefont
  {S.}~\bibnamefont {Valkiers}}, \bibinfo {author} {\bibfnamefont
  {A.}~\bibnamefont {Waseda}}, \ and\ \bibinfo {author} {\bibfnamefont
  {S.}~\bibnamefont {Zakel}},\ }\href@noop {} {\bibfield  {journal} {\bibinfo
  {journal} {Metrologia}\ }\textbf {\bibinfo {volume} {48}},\ \bibinfo {pages}
  {S1} (\bibinfo {year} {2011})}\BibitemShut {NoStop}%
\bibitem [{\citenamefont {Azuma}\ \emph {et~al.}(2015)\citenamefont {Azuma},
  \citenamefont {Barat}, \citenamefont {Bartl}, \citenamefont {Bettin},
  \citenamefont {Borys}, \citenamefont {Busch}, \citenamefont {Cibik},
  \citenamefont {D’Agostino}, \citenamefont {Fujii}, \citenamefont
  {Fujimoto}, \citenamefont {Hioki}, \citenamefont {Krumrey}, \citenamefont
  {Kuetgens}, \citenamefont {Kuramoto}, \citenamefont {Mana}, \citenamefont
  {Massa}, \citenamefont {Mee\ss}, \citenamefont {Mizushima}, \citenamefont
  {Narukawa}, \citenamefont {Nicolaus}, \citenamefont {Pramann}, \citenamefont
  {Rabb}, \citenamefont {Rienitz}, \citenamefont {Sasso}, \citenamefont
  {Stock}, \citenamefont {Vocke}, \citenamefont {Waseda}, \citenamefont
  {Wundrack},\ and\ \citenamefont {Zakel}}]{Azuma:2015}%
  \BibitemOpen
  \bibfield  {author} {\bibinfo {author} {\bibfnamefont {Y.}~\bibnamefont
  {Azuma}}, \bibinfo {author} {\bibfnamefont {P.}~\bibnamefont {Barat}},
  \bibinfo {author} {\bibfnamefont {G.}~\bibnamefont {Bartl}}, \bibinfo
  {author} {\bibfnamefont {H.}~\bibnamefont {Bettin}}, \bibinfo {author}
  {\bibfnamefont {M.}~\bibnamefont {Borys}}, \bibinfo {author} {\bibfnamefont
  {I.}~\bibnamefont {Busch}}, \bibinfo {author} {\bibfnamefont
  {L.}~\bibnamefont {Cibik}}, \bibinfo {author} {\bibfnamefont
  {G.}~\bibnamefont {D’Agostino}}, \bibinfo {author} {\bibfnamefont
  {K.}~\bibnamefont {Fujii}}, \bibinfo {author} {\bibfnamefont
  {H.}~\bibnamefont {Fujimoto}}, \bibinfo {author} {\bibfnamefont
  {A.}~\bibnamefont {Hioki}}, \bibinfo {author} {\bibfnamefont
  {M.}~\bibnamefont {Krumrey}}, \bibinfo {author} {\bibfnamefont
  {U.}~\bibnamefont {Kuetgens}}, \bibinfo {author} {\bibfnamefont
  {N.}~\bibnamefont {Kuramoto}}, \bibinfo {author} {\bibfnamefont
  {G.}~\bibnamefont {Mana}}, \bibinfo {author} {\bibfnamefont {E.}~\bibnamefont
  {Massa}}, \bibinfo {author} {\bibfnamefont {R.}~\bibnamefont {Mee\ss}},
  \bibinfo {author} {\bibfnamefont {S.}~\bibnamefont {Mizushima}}, \bibinfo
  {author} {\bibfnamefont {T.}~\bibnamefont {Narukawa}}, \bibinfo {author}
  {\bibfnamefont {A.}~\bibnamefont {Nicolaus}}, \bibinfo {author}
  {\bibfnamefont {A.}~\bibnamefont {Pramann}}, \bibinfo {author} {\bibfnamefont
  {S.~A.}\ \bibnamefont {Rabb}}, \bibinfo {author} {\bibfnamefont
  {O.}~\bibnamefont {Rienitz}}, \bibinfo {author} {\bibfnamefont
  {C.}~\bibnamefont {Sasso}}, \bibinfo {author} {\bibfnamefont
  {M.}~\bibnamefont {Stock}}, \bibinfo {author} {\bibfnamefont {R.~D.~J.}\
  \bibnamefont {Vocke}}, \bibinfo {author} {\bibfnamefont {A.}~\bibnamefont
  {Waseda}}, \bibinfo {author} {\bibfnamefont {S.}~\bibnamefont {Wundrack}}, \
  and\ \bibinfo {author} {\bibfnamefont {S.}~\bibnamefont {Zakel}},\
  }\href@noop {} {\bibfield  {journal} {\bibinfo  {journal} {Metrologia}\
  }\textbf {\bibinfo {volume} {52}},\ \bibinfo {pages} {360} (\bibinfo {year}
  {2015})}\BibitemShut {NoStop}%
\bibitem [{\citenamefont {Massa}\ and\ \citenamefont
  {Nicolaus}(2011)}]{Massa:2011f}%
  \BibitemOpen
  \bibfield  {author} {\bibinfo {author} {\bibfnamefont {E.}~\bibnamefont
  {Massa}}\ and\ \bibinfo {author} {\bibfnamefont {A.}~\bibnamefont
  {Nicolaus}},\ }\href {http://stacks.iop.org/0026-1394/48/i=2/a=E01}
  {\bibfield  {journal} {\bibinfo  {journal} {Metrologia}\ }\textbf {\bibinfo
  {volume} {48}},\ \bibinfo {pages} {E1} (\bibinfo {year} {2011})}\BibitemShut
  {NoStop}%
\bibitem [{\citenamefont {Nicolaus}\ \emph {et~al.}(2013)\citenamefont
  {Nicolaus}, \citenamefont {Bartl}, \citenamefont {Bettin}, \citenamefont
  {Borys}, \citenamefont {Firlus}, \citenamefont {Busch}, \citenamefont
  {Felgner}, \citenamefont {Kruger-Sehm}, \citenamefont {Krumrey},
  \citenamefont {Krystek},\ and\ \citenamefont {Kuetgens}}]{Nicolaus:2013}%
  \BibitemOpen
  \bibfield  {author} {\bibinfo {author} {\bibfnamefont {R.}~\bibnamefont
  {Nicolaus}}, \bibinfo {author} {\bibfnamefont {G.}~\bibnamefont {Bartl}},
  \bibinfo {author} {\bibfnamefont {H.}~\bibnamefont {Bettin}}, \bibinfo
  {author} {\bibfnamefont {M.}~\bibnamefont {Borys}}, \bibinfo {author}
  {\bibfnamefont {M.}~\bibnamefont {Firlus}}, \bibinfo {author} {\bibfnamefont
  {I.}~\bibnamefont {Busch}}, \bibinfo {author} {\bibfnamefont
  {A.}~\bibnamefont {Felgner}}, \bibinfo {author} {\bibfnamefont
  {R.}~\bibnamefont {Kruger-Sehm}}, \bibinfo {author} {\bibfnamefont
  {M.}~\bibnamefont {Krumrey}}, \bibinfo {author} {\bibfnamefont
  {M.}~\bibnamefont {Krystek}}, \ and\ \bibinfo {author} {\bibfnamefont
  {U.}~\bibnamefont {Kuetgens}},\ }\href {\doibase 10.1109/TIM.2013.2242633}
  {\bibfield  {journal} {\bibinfo  {journal} {Instrumentation and Measurement,
  IEEE Transactions on}\ }\textbf {\bibinfo {volume} {62}},\ \bibinfo {pages}
  {1499} (\bibinfo {year} {2013})}\BibitemShut {NoStop}%
\bibitem [{\citenamefont {Mohr}, \citenamefont {Taylor},\ and\ \citenamefont
  {Newell}(2012{\natexlab{a}})}]{CODATA:2012a}%
  \BibitemOpen
  \bibfield  {author} {\bibinfo {author} {\bibfnamefont {P.~J.}\ \bibnamefont
  {Mohr}}, \bibinfo {author} {\bibfnamefont {B.~N.}\ \bibnamefont {Taylor}}, \
  and\ \bibinfo {author} {\bibfnamefont {D.~B.}\ \bibnamefont {Newell}},\
  }\href {\doibase 10.1103/RevModPhys.84.1527} {\bibfield  {journal} {\bibinfo
  {journal} {Rev. Mod. Phys.}\ }\textbf {\bibinfo {volume} {84}},\ \bibinfo
  {pages} {1527} (\bibinfo {year} {2012}{\natexlab{a}})}\BibitemShut {NoStop}%
\bibitem [{\citenamefont {Mohr}, \citenamefont {Taylor},\ and\ \citenamefont
  {Newell}(2012{\natexlab{b}})}]{CODATA:2012b}%
  \BibitemOpen
  \bibfield  {author} {\bibinfo {author} {\bibfnamefont {P.~J.}\ \bibnamefont
  {Mohr}}, \bibinfo {author} {\bibfnamefont {B.~N.}\ \bibnamefont {Taylor}}, \
  and\ \bibinfo {author} {\bibfnamefont {D.~B.}\ \bibnamefont {Newell}},\
  }\href {\doibase http://dx.doi.org/10.1063/1.4724320} {\bibfield  {journal}
  {\bibinfo  {journal} {Journal of Physical and Chemical Reference Data}\
  }\textbf {\bibinfo {volume} {41}},\ \bibinfo {eid} {043109} (\bibinfo {year}
  {2012}{\natexlab{b}})}\BibitemShut {NoStop}%
\bibitem [{\citenamefont {Massa}\ \emph
  {et~al.}(2011{\natexlab{a}})\citenamefont {Massa}, \citenamefont {Mana},
  \citenamefont {Kuetgens},\ and\ \citenamefont {Ferroglio}}]{Massa:2011a}%
  \BibitemOpen
  \bibfield  {author} {\bibinfo {author} {\bibfnamefont {E.}~\bibnamefont
  {Massa}}, \bibinfo {author} {\bibfnamefont {G.}~\bibnamefont {Mana}},
  \bibinfo {author} {\bibfnamefont {U.}~\bibnamefont {Kuetgens}}, \ and\
  \bibinfo {author} {\bibfnamefont {L.}~\bibnamefont {Ferroglio}},\ }\href@noop
  {} {\bibfield  {journal} {\bibinfo  {journal} {Metrologia}\ }\textbf
  {\bibinfo {volume} {48}},\ \bibinfo {pages} {S37} (\bibinfo {year}
  {2011}{\natexlab{a}})}\BibitemShut {NoStop}%
\bibitem [{\citenamefont {Massa}\ \emph {et~al.}(2015)\citenamefont {Massa},
  \citenamefont {Sasso}, \citenamefont {Mana},\ and\ \citenamefont
  {Palmisano}}]{Massa:2015}%
  \BibitemOpen
  \bibfield  {author} {\bibinfo {author} {\bibfnamefont {E.}~\bibnamefont
  {Massa}}, \bibinfo {author} {\bibfnamefont {C.}~\bibnamefont {Sasso}},
  \bibinfo {author} {\bibfnamefont {G.}~\bibnamefont {Mana}}, \ and\ \bibinfo
  {author} {\bibfnamefont {C.}~\bibnamefont {Palmisano}},\ }\href@noop {}
  {\bibfield  {journal} {\bibinfo  {journal} {J.\ Phys.\ Chem.\ Ref.\ Data}\
  }\textbf {\bibinfo {volume} {44}},\ \bibinfo {pages} {accepted} (\bibinfo
  {year} {2015})}\BibitemShut {NoStop}%
\bibitem [{\citenamefont {Pramann}\ \emph {et~al.}(2011)\citenamefont
  {Pramann}, \citenamefont {Rienitz}, \citenamefont {Schiel}, \citenamefont
  {Schlote}, \citenamefont {G\"uttler},\ and\ \citenamefont
  {Valkiers}}]{Pramann:2011}%
  \BibitemOpen
  \bibfield  {author} {\bibinfo {author} {\bibfnamefont {A.}~\bibnamefont
  {Pramann}}, \bibinfo {author} {\bibfnamefont {O.}~\bibnamefont {Rienitz}},
  \bibinfo {author} {\bibfnamefont {D.}~\bibnamefont {Schiel}}, \bibinfo
  {author} {\bibfnamefont {J.}~\bibnamefont {Schlote}}, \bibinfo {author}
  {\bibfnamefont {B.}~\bibnamefont {G\"uttler}}, \ and\ \bibinfo {author}
  {\bibfnamefont {S.}~\bibnamefont {Valkiers}},\ }\href@noop {} {\bibfield
  {journal} {\bibinfo  {journal} {Metrologia}\ }\textbf {\bibinfo {volume}
  {48}},\ \bibinfo {pages} {S20} (\bibinfo {year} {2011})}\BibitemShut
  {NoStop}%
\bibitem [{\citenamefont {Narukawa}\ \emph {et~al.}(2014)\citenamefont
  {Narukawa}, \citenamefont {Hioki}, \citenamefont {Kuramoto},\ and\
  \citenamefont {Fujii}}]{Narukawa:2014}%
  \BibitemOpen
  \bibfield  {author} {\bibinfo {author} {\bibfnamefont {T.}~\bibnamefont
  {Narukawa}}, \bibinfo {author} {\bibfnamefont {A.}~\bibnamefont {Hioki}},
  \bibinfo {author} {\bibfnamefont {N.}~\bibnamefont {Kuramoto}}, \ and\
  \bibinfo {author} {\bibfnamefont {K.}~\bibnamefont {Fujii}},\ }\href@noop {}
  {\bibfield  {journal} {\bibinfo  {journal} {Metrologia}\ }\textbf {\bibinfo
  {volume} {51}},\ \bibinfo {pages} {161} (\bibinfo {year} {2014})}\BibitemShut
  {NoStop}%
\bibitem [{\citenamefont {Jr}, \citenamefont {Rabb},\ and\ \citenamefont
  {Turk}(2014)}]{Vocke:2014}%
  \BibitemOpen
  \bibfield  {author} {\bibinfo {author} {\bibfnamefont {R.~D.~V.}\
  \bibnamefont {Jr}}, \bibinfo {author} {\bibfnamefont {S.~A.}\ \bibnamefont
  {Rabb}}, \ and\ \bibinfo {author} {\bibfnamefont {G.~C.}\ \bibnamefont
  {Turk}},\ }\href@noop {} {\bibfield  {journal} {\bibinfo  {journal}
  {Metrologia}\ }\textbf {\bibinfo {volume} {51}},\ \bibinfo {pages} {361}
  (\bibinfo {year} {2014})}\BibitemShut {NoStop}%
\bibitem [{\citenamefont {Pramann}\ \emph {et~al.}(2014)\citenamefont
  {Pramann}, \citenamefont {Rienitz}, \citenamefont {Noordmann}, \citenamefont
  {G\"uttler},\ and\ \citenamefont {Schiel}}]{Pramann:2014}%
  \BibitemOpen
  \bibfield  {author} {\bibinfo {author} {\bibfnamefont {A.}~\bibnamefont
  {Pramann}}, \bibinfo {author} {\bibfnamefont {O.}~\bibnamefont {Rienitz}},
  \bibinfo {author} {\bibfnamefont {J.}~\bibnamefont {Noordmann}}, \bibinfo
  {author} {\bibfnamefont {B.}~\bibnamefont {G\"uttler}}, \ and\ \bibinfo
  {author} {\bibfnamefont {D.}~\bibnamefont {Schiel}},\ }\href@noop {}
  {\bibfield  {journal} {\bibinfo  {journal} {Zeitschrift f\"ur Physikalische
  Chemie}\ }\textbf {\bibinfo {volume} {228}},\ \bibinfo {pages} {405}
  (\bibinfo {year} {2014})}\BibitemShut {NoStop}%
\bibitem [{\citenamefont {Picard}\ \emph {et~al.}(2011)\citenamefont {Picard},
  \citenamefont {Barat}, \citenamefont {Borys}, \citenamefont {Firlus},\ and\
  \citenamefont {Mizushima}}]{Picard:2011}%
  \BibitemOpen
  \bibfield  {author} {\bibinfo {author} {\bibfnamefont {A.}~\bibnamefont
  {Picard}}, \bibinfo {author} {\bibfnamefont {P.}~\bibnamefont {Barat}},
  \bibinfo {author} {\bibfnamefont {M.}~\bibnamefont {Borys}}, \bibinfo
  {author} {\bibfnamefont {M.}~\bibnamefont {Firlus}}, \ and\ \bibinfo {author}
  {\bibfnamefont {S.}~\bibnamefont {Mizushima}},\ }\href@noop {} {\bibfield
  {journal} {\bibinfo  {journal} {Metrologia}\ }\textbf {\bibinfo {volume}
  {48}},\ \bibinfo {pages} {S112} (\bibinfo {year} {2011})}\BibitemShut
  {NoStop}%
\bibitem [{\citenamefont {Kuramoto}, \citenamefont {Fujii},\ and\ \citenamefont
  {Yamazawa}(2011)}]{Kuramoto:2011}%
  \BibitemOpen
  \bibfield  {author} {\bibinfo {author} {\bibfnamefont {N.}~\bibnamefont
  {Kuramoto}}, \bibinfo {author} {\bibfnamefont {K.}~\bibnamefont {Fujii}}, \
  and\ \bibinfo {author} {\bibfnamefont {K.}~\bibnamefont {Yamazawa}},\
  }\href@noop {} {\bibfield  {journal} {\bibinfo  {journal} {Metrologia}\
  }\textbf {\bibinfo {volume} {48}},\ \bibinfo {pages} {S83} (\bibinfo {year}
  {2011})}\BibitemShut {NoStop}%
\bibitem [{\citenamefont {Bartl}\ \emph {et~al.}(2011)\citenamefont {Bartl},
  \citenamefont {Bettin}, \citenamefont {Krystek}, \citenamefont {Mai},
  \citenamefont {Nicolaus},\ and\ \citenamefont {Peter}}]{Bartl:2011}%
  \BibitemOpen
  \bibfield  {author} {\bibinfo {author} {\bibfnamefont {G.}~\bibnamefont
  {Bartl}}, \bibinfo {author} {\bibfnamefont {H.}~\bibnamefont {Bettin}},
  \bibinfo {author} {\bibfnamefont {M.}~\bibnamefont {Krystek}}, \bibinfo
  {author} {\bibfnamefont {T.}~\bibnamefont {Mai}}, \bibinfo {author}
  {\bibfnamefont {A.}~\bibnamefont {Nicolaus}}, \ and\ \bibinfo {author}
  {\bibfnamefont {A.}~\bibnamefont {Peter}},\ }\href@noop {} {\bibfield
  {journal} {\bibinfo  {journal} {Metrologia}\ }\textbf {\bibinfo {volume}
  {48}},\ \bibinfo {pages} {S96} (\bibinfo {year} {2011})}\BibitemShut
  {NoStop}%
\bibitem [{\citenamefont {Fujimoto}, \citenamefont {Waseda},\ and\
  \citenamefont {Zhang}(2011)}]{Fujimoto:2011}%
  \BibitemOpen
  \bibfield  {author} {\bibinfo {author} {\bibfnamefont {H.}~\bibnamefont
  {Fujimoto}}, \bibinfo {author} {\bibfnamefont {A.}~\bibnamefont {Waseda}}, \
  and\ \bibinfo {author} {\bibfnamefont {X.~W.}\ \bibnamefont {Zhang}},\
  }\href@noop {} {\bibfield  {journal} {\bibinfo  {journal} {Metrologia}\
  }\textbf {\bibinfo {volume} {48}},\ \bibinfo {pages} {S55} (\bibinfo {year}
  {2011})}\BibitemShut {NoStop}%
\bibitem [{\citenamefont {Massa}\ \emph
  {et~al.}(2011{\natexlab{b}})\citenamefont {Massa}, \citenamefont {Mana},
  \citenamefont {Ferroglio}, \citenamefont {Kessler}, \citenamefont {Schiel},\
  and\ \citenamefont {Zakel}}]{Massa:2011b}%
  \BibitemOpen
  \bibfield  {author} {\bibinfo {author} {\bibfnamefont {E.}~\bibnamefont
  {Massa}}, \bibinfo {author} {\bibfnamefont {G.}~\bibnamefont {Mana}},
  \bibinfo {author} {\bibfnamefont {L.}~\bibnamefont {Ferroglio}}, \bibinfo
  {author} {\bibfnamefont {E.~G.}\ \bibnamefont {Kessler}}, \bibinfo {author}
  {\bibfnamefont {D.}~\bibnamefont {Schiel}}, \ and\ \bibinfo {author}
  {\bibfnamefont {S.}~\bibnamefont {Zakel}},\ }\href@noop {} {\bibfield
  {journal} {\bibinfo  {journal} {Metrologia}\ }\textbf {\bibinfo {volume}
  {48}},\ \bibinfo {pages} {S44} (\bibinfo {year}
  {2011}{\natexlab{b}})}\BibitemShut {NoStop}%
\bibitem [{\citenamefont {Zakel}\ \emph {et~al.}(2011)\citenamefont {Zakel},
  \citenamefont {Wundrack}, \citenamefont {Niemann}, \citenamefont {Rienitz},\
  and\ \citenamefont {Schiel}}]{Zakel:2011}%
  \BibitemOpen
  \bibfield  {author} {\bibinfo {author} {\bibfnamefont {S.}~\bibnamefont
  {Zakel}}, \bibinfo {author} {\bibfnamefont {S.}~\bibnamefont {Wundrack}},
  \bibinfo {author} {\bibfnamefont {H.}~\bibnamefont {Niemann}}, \bibinfo
  {author} {\bibfnamefont {O.}~\bibnamefont {Rienitz}}, \ and\ \bibinfo
  {author} {\bibfnamefont {D.}~\bibnamefont {Schiel}},\ }\href@noop {}
  {\bibfield  {journal} {\bibinfo  {journal} {Metrologia}\ }\textbf {\bibinfo
  {volume} {48}},\ \bibinfo {pages} {S14} (\bibinfo {year} {2011})}\BibitemShut
  {NoStop}%
\bibitem [{\citenamefont {Melis}, \citenamefont {Colombo},\ and\ \citenamefont
  {Mana}(2015)}]{Melis:2015}%
  \BibitemOpen
  \bibfield  {author} {\bibinfo {author} {\bibfnamefont {C.}~\bibnamefont
  {Melis}}, \bibinfo {author} {\bibfnamefont {L.}~\bibnamefont {Colombo}}, \
  and\ \bibinfo {author} {\bibfnamefont {G.}~\bibnamefont {Mana}},\ }\href
  {http://stacks.iop.org/0026-1394/52/i=2/a=214} {\bibfield  {journal}
  {\bibinfo  {journal} {Metrologia}\ }\textbf {\bibinfo {volume} {52}},\
  \bibinfo {pages} {214} (\bibinfo {year} {2015})}\BibitemShut {NoStop}%
\bibitem [{\citenamefont {Busch}\ \emph {et~al.}(2011)\citenamefont {Busch},
  \citenamefont {Azuma}, \citenamefont {Bettin}, \citenamefont {Cibik},
  \citenamefont {Fuchs}, \citenamefont {Fujii}, \citenamefont {Krumrey},
  \citenamefont {Kuetgens}, \citenamefont {Kuramoto},\ and\ \citenamefont
  {Mizushima}}]{Busch:2011}%
  \BibitemOpen
  \bibfield  {author} {\bibinfo {author} {\bibfnamefont {I.}~\bibnamefont
  {Busch}}, \bibinfo {author} {\bibfnamefont {Y.}~\bibnamefont {Azuma}},
  \bibinfo {author} {\bibfnamefont {H.}~\bibnamefont {Bettin}}, \bibinfo
  {author} {\bibfnamefont {L.}~\bibnamefont {Cibik}}, \bibinfo {author}
  {\bibfnamefont {P.}~\bibnamefont {Fuchs}}, \bibinfo {author} {\bibfnamefont
  {K.}~\bibnamefont {Fujii}}, \bibinfo {author} {\bibfnamefont
  {M.}~\bibnamefont {Krumrey}}, \bibinfo {author} {\bibfnamefont
  {U.}~\bibnamefont {Kuetgens}}, \bibinfo {author} {\bibfnamefont
  {N.}~\bibnamefont {Kuramoto}}, \ and\ \bibinfo {author} {\bibfnamefont
  {S.}~\bibnamefont {Mizushima}},\ }\href@noop {} {\bibfield  {journal}
  {\bibinfo  {journal} {Metrologia}\ }\textbf {\bibinfo {volume} {48}},\
  \bibinfo {pages} {S62} (\bibinfo {year} {2011})}\BibitemShut {NoStop}%
\bibitem [{\citenamefont {Mana}, \citenamefont {Rienitz},\ and\ \citenamefont
  {Pramann}(2010)}]{Mana:2010}%
  \BibitemOpen
  \bibfield  {author} {\bibinfo {author} {\bibfnamefont {G.}~\bibnamefont
  {Mana}}, \bibinfo {author} {\bibfnamefont {O.}~\bibnamefont {Rienitz}}, \
  and\ \bibinfo {author} {\bibfnamefont {A.}~\bibnamefont {Pramann}},\ }\href
  {http://stacks.iop.org/0026-1394/47/i=4/a=012} {\bibfield  {journal}
  {\bibinfo  {journal} {Metrologia}\ }\textbf {\bibinfo {volume} {47}},\
  \bibinfo {pages} {460} (\bibinfo {year} {2010})}\BibitemShut {NoStop}%
\bibitem [{\citenamefont {D’Agostino}\ \emph {et~al.}(2015)\citenamefont
  {D’Agostino}, \citenamefont {Mana}, \citenamefont {Oddone}, \citenamefont
  {Prata}, \citenamefont {Di~Luzio},\ and\ \citenamefont
  {Pramann}}]{D'Agostino:2015}%
  \BibitemOpen
  \bibfield  {author} {\bibinfo {author} {\bibfnamefont {G.}~\bibnamefont
  {D’Agostino}}, \bibinfo {author} {\bibfnamefont {G.}~\bibnamefont {Mana}},
  \bibinfo {author} {\bibfnamefont {M.}~\bibnamefont {Oddone}}, \bibinfo
  {author} {\bibfnamefont {M.}~\bibnamefont {Prata}}, \bibinfo {author}
  {\bibfnamefont {M.}~\bibnamefont {Di~Luzio}}, \ and\ \bibinfo {author}
  {\bibfnamefont {A.}~\bibnamefont {Pramann}},\ }\href@noop {} {\bibfield
  {journal} {\bibinfo  {journal} {Anal.\ Chem.}\ ,\ \bibinfo {pages}
  {submitted}} (\bibinfo {year} {2015})}\BibitemShut {NoStop}%
\bibitem [{\citenamefont {Quagliotti}\ \emph {et~al.}(2013)\citenamefont
  {Quagliotti}, \citenamefont {Mana}, \citenamefont {Massa}, \citenamefont
  {Sasso},\ and\ \citenamefont {Kuetgens}}]{Quagliotti:2013}%
  \BibitemOpen
  \bibfield  {author} {\bibinfo {author} {\bibfnamefont {D.}~\bibnamefont
  {Quagliotti}}, \bibinfo {author} {\bibfnamefont {G.}~\bibnamefont {Mana}},
  \bibinfo {author} {\bibfnamefont {E.}~\bibnamefont {Massa}}, \bibinfo
  {author} {\bibfnamefont {C.}~\bibnamefont {Sasso}}, \ and\ \bibinfo {author}
  {\bibfnamefont {U.}~\bibnamefont {Kuetgens}},\ }\href@noop {} {\bibfield
  {journal} {\bibinfo  {journal} {Metrologia}\ }\textbf {\bibinfo {volume}
  {50}},\ \bibinfo {pages} {243} (\bibinfo {year} {2013})}\BibitemShut
  {NoStop}%
\bibitem [{\citenamefont {Stock}\ \emph {et~al.}(2015)\citenamefont {Stock},
  \citenamefont {Barat}, \citenamefont {Davis}, \citenamefont {Picard},\ and\
  \citenamefont {Milton}}]{Stock:2015}%
  \BibitemOpen
  \bibfield  {author} {\bibinfo {author} {\bibfnamefont {M.}~\bibnamefont
  {Stock}}, \bibinfo {author} {\bibfnamefont {P.}~\bibnamefont {Barat}},
  \bibinfo {author} {\bibfnamefont {R.~S.}\ \bibnamefont {Davis}}, \bibinfo
  {author} {\bibfnamefont {A.}~\bibnamefont {Picard}}, \ and\ \bibinfo {author}
  {\bibfnamefont {M.~J.~T.}\ \bibnamefont {Milton}},\ }\href
  {http://stacks.iop.org/0026-1394/52/i=2/a=310} {\bibfield  {journal}
  {\bibinfo  {journal} {Metrologia}\ }\textbf {\bibinfo {volume} {52}},\
  \bibinfo {pages} {310} (\bibinfo {year} {2015})}\BibitemShut {NoStop}%
\bibitem [{\citenamefont {Mizushima}(2004)}]{Mizushima:2004}%
  \BibitemOpen
  \bibfield  {author} {\bibinfo {author} {\bibfnamefont {S.}~\bibnamefont
  {Mizushima}},\ }\href {http://stacks.iop.org/0026-1394/41/i=3/a=005}
  {\bibfield  {journal} {\bibinfo  {journal} {Metrologia}\ }\textbf {\bibinfo
  {volume} {41}},\ \bibinfo {pages} {137} (\bibinfo {year} {2004})}\BibitemShut
  {NoStop}%
\end{thebibliography}%

\end{document}